\documentclass{amsart}

\usepackage{times}
\usepackage{amsmath}
\usepackage{amssymb}
\usepackage{amsthm}
\usepackage{mathabx}
\usepackage[latin1]{inputenc}
\usepackage{eurosym}
\usepackage{graphicx}
\usepackage{epstopdf} 
\usepackage{float}
\usepackage{caption}
\usepackage{subcaption}
\usepackage{epsfig}
\usepackage{color}
\usepackage{algorithm}
\usepackage{algorithmic}
\usepackage{enumerate}
\usepackage{url}

\begin{document}

\title{Recommendation via matrix completion using Kolmogorov complexity}

\author{Guilherme Ramos$^\ast$}
\address{Department of Mathematics, Instituto Superior T\'ecnico,
University of Lisbon, Lisbon, Portugal}
\curraddr{Instituto de Telecomunica\c{c}\~oes, Instituto Superior T\'ecnico,
University of Lisbon, Lisbon, Portugal}
\email{guilherme.ramos@tecnico.ulisboa.pt}
\thanks{This work was developed under the scope of R\&D Unit 50008, financed by the applicable financial framework (FCT/MEC through national funds and when applicable co-funded by FEDER - PT2020 partnership agreement).
The second author acknowledges the support of the DP-PMI and Funda\c{c}\~ao para a Ci\^encia e a Tecnologia (Portugal), namely through scholarship SFRH/BD/52242/2013.}

\author{Jo\~{a}o Sa\'{u}de$^\ast$}
\address{Department of Electrical and Computer Engineering, Carnegie Mellon University, Pittsburgh, PA 15213}
\curraddr{LARSyS, Instituto Superior T\'ecnico,
University of Lisbon, Lisbon, Portugal}
\email{jsaude@andrew.cmu.edu}
\thanks{$^\ast$ The first two authors contributed equally to this work.\\
The work was partially supported through the Carnegie Mellon/Portugal Program managed by ICTI from FCT and by FCT grant SFRH/BD/52162/2013.}

\author{Carlos Caleiro}
\address{Department of Mathematics, Instituto Superior T\'ecnico,
University of Lisbon, Lisbon, Portugal}
\email{guilherme.ramos@tecnico.ulisboa.pt}
\thanks{}

\author{Soummya Kar}
\address{Department of Electrical and Computer Engineering, Carnegie Mellon University, Pittsburgh, PA 15213}
\email{soummyak@andrew.cmu.edu}
\thanks{}


\date{April 3, 2017 and, in revised form, July 12, 2017.}



\begin{abstract}
	A usual way to model a recommendation system is as a matrix completion problem.
	There are several matrix completion methods, typically using optimization approaches or collaborative filtering.
	Most approaches assume that the matrix is either low rank, or that there are a small number of latent variables that encode the full problem.
	Here, we propose a novel matrix completion algorithm for recommendation systems, without any assumptions on the rank and that is model free, i.e., the entries are not assumed to be a function of some latent variables.
	Instead, we use a technique akin to information theory.
	Our method performs hybrid neighborhood-based collaborative filtering using Kolmogorov complexity.
	It decouples the matrix completion into a vector completion problem for each user.
	The recommendation for one user is thus independent of the recommendation for other users.
	This makes the algorithm scalable because the computations are highly parallelizable.
	Our results are competitive with state-of-the-art approaches on both synthetic and real-world dataset benchmarks.	
\end{abstract}

\maketitle

\section{Introduction}
	The continuing increase of online services, like e-commerce, audio/video streaming, online news, reviews and opinion providers, potentiate the demand for recommendation of online services/products.
	The huge amount of services/products available makes the choice a difficult matter.
	Users rely not only on reviews and ratings, but also take into account automatic suggestions by the providers.
	Therefore, automatic recommendation systems became essential and widely used by providers and consumers.
	
	\textbf{Previous work.} Several approaches to the matrix completion problem reformulate it into an optimization problem, assuming that the matrix to recover has low rank, and that the observed entries' positions are sampled from accordingly to a uniform distribution, see \cite{candes2010power}. 
	Although the rank minimization problem is NP-hard, approaches following the ideas in \cite{candes2010power} are used with relative success. It consists in relaxing the problem so that it becomes convex, and then in minimizing the nuclear norm of the matrix. These methods are very used in practice. 
	In other approaches, it is assumed that the matrix to complete is high rank.
	This also entails dealing with a NP-hard problem.
	Nonetheless, under certain assumptions, some incomplete high rank or even full rank matrix can be completed, as in \cite{balzano2012high}.
	In their work, the authors assume that the columns of the matrix to complete belong to a union of multiple low-rank subspaces. This way, the problem can be viewed as a missing-data version of the subspace clustering problem.
	
	Collaborative filtering approaches are mainly divided in two research lines: model-based and neighborhood-based \cite{ricci2011introduction}.
	The first line tries to model latent factors of both users and items and is widely used due to its demonstrated success for movie recommendation in the Netflix prize \cite{bennett2007netflix}.  
	The second line of research does recommendation based on users with similar tastes/preferences or items that are similar to the users preferences.
	This last line further divides into three main approaches, user-based, item-based and hybrid.
	In user-based methods, we select a set of similar users based on similarity among them to recommend items as, for example, in \cite{zhao2010user}.
	Item-based methods, are analogous, but performed using similarities among the items as, for instance, in \cite{sarwar2001item}.
	The hybrid approaches combine the previous, see \cite{wang2006unifying}.
	In this work we use hybrid collaborative filtering to address the matrix completion problem.

   	In recent work by \cite{ganti2015matrix}, the authors addressed the matrix completion problem not assuming that the matrix is low rank, as is most common.
	They consider the case when entries of a low-rank matrix are recovered through a Lipschitz monotonic function, transforming the matrix into a high rank one, and the aim is to recover the unobserved entries.
	For the task, they propose an iterative method that alternates between estimating a low rank matrix, and estimating the monotonic function, in order to recover the missing elements of the high rank matrix.
	 Further, they provide Mean Square Error (MSE) bounds for the recover error, based on the rank of the matrix, its size, and properties of the nonlinear transformation.
	 The algorithm only applies to functions that are nonlinear monotonic transformations of the inner product of latent features.
	 
	  In \cite{song2016blind}, the authors address the matrix completion problem using a novel framework for nonparametric regression over latent variable models. 
 	They propose to model the unknown matrix entries as a Lipschitz function of two latent variables, one for users and another for items. 
 	Using the Taylor expansion of the unknown function, around different points, they can define the value of the missing entry as a weighted convex combination of the known entries.
 	They use as measure of similarity the sample variance between rows and columns.
 	Then, they use kernel regression to perform local smoothing.

	In \cite{wang2006unifying}, the authors present a generative probabilistic framework that considers similarity between users and between items. The prediction of each unknown matrix entry is made by averaging the individual ratings weighted by the users confidence. This allows the authors to take advantage of both user correlations and item correlations to better estimate the missing entries of the rating matrix. The authors consider three similarity matrices in their work.

\textbf{Main contributions.} We present a simple approach to build a recommendation system based on matrix completion by performing hybrid (user and item) neighborhood-based collaborative-filtering, summarized in Algorithm~\ref{MatComp} from Section~\ref{Main}. Our method explores Kolmogorov complexity to construct a similarity measure from information theory \cite{cover2012elements}, and to propose new similarity measures.
	The algorithm that we propose is modular and the recommendation for each user can be computed independently. 
	Further, our algorithms works with a small number of data points, it works for both low-rank and high-rank matrix completion, without the need of any initialization.
	Last, the computations of the algorithm can be done in a distributed fashion, making it scalable. 
	
\textbf{Paper structure.} The remainder of the paper is organized as follows. In Section~\ref{Setup}, we introduce some notation and present our setup specification. In Section~\ref{Experimentos}, we use our matrix completion algorithm, Algorithm~\ref{MatComp}, to evaluate its performance, with both synthetic data and real-world datasets. Section~\ref{Conclusion} concludes the paper and draws avenues for further research. 
\section{Setup}\label{Setup}
We first introduce some notation to make the paper self-contained, and then we present our matrix completion algorithm and its computational complexity analysis. 
\subsection{Notation}\label{Notation}
We denote the set of $n$ users by $\mathcal U=\{u_1,\hdots,u_n\}$, the set of $m$ items by $\mathcal I=\{o_1,\hdots,o_m\}$, and the $n\times m$ matrix of ratings by $M$, where $M_{uo}$ denotes the rating that user $u$ gave to item $o$. The entries take values on the allowed ratings together with a special number denoting the absence of rating (in this work this value is $0$).
We adopt standard notation to denote matrices and vectors. For a matrix $M$, we denote the $i$th row of $M$ by $M_{i\cdot}$, the $j$th column of $M$ by $M_{\cdot j}$, and the $j$th column of the $i$th row by $M_{ij}$.
Given a set of objects $\mathcal X$, a \emph{similarity} is a function $s:\mathcal X\times\mathcal X\to [0,1]$ such that whenever $x\in \mathcal X$, $s(x,x)=1$.
For a square matrix representing similarities we use the letter $S$ indexed by $\mathcal U$ or $\mathcal I$, if the similarity matrix represents similarities between users or items, respectively.
Further, given two vectors with dimension $n$, $x$ and $y$, we denote by $x\odot y$ the vector whose entries are the product of the entries of $x$ and $y$, i.e., $x\odot y=(x_1y_1,\hdots,x_ny_n)$.
Finally, we use the semi-norm $\|\cdot\|_0$. Given a vector $x$, $\|x\|_0$ is the number of non zero entries of $x$.

\subsection{Setup specification}\label{Main}
\begin{figure}
		\centering
 		\includegraphics[width=0.6\textwidth]{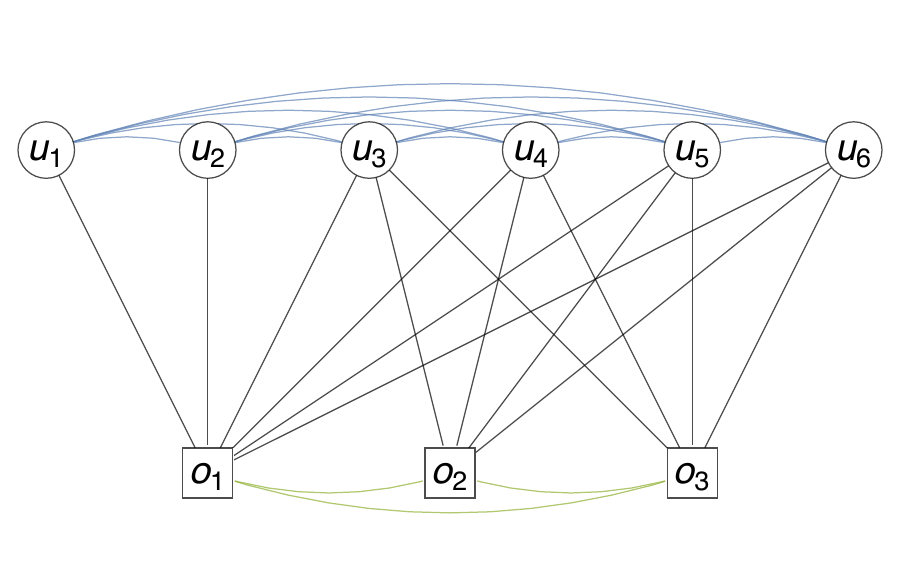}
	 	\caption{Graph representing $N$ users, $u_i$, $M$ items, $o_j$. The black edges between users and items represent the products each user rated. 
				 The blue edges (between users) represent the weights computed in the matrix $ S_{\mathcal U}$. The green edges (between items) represent the weights computed in the matrix $ S_{\mathcal I}$.}
	 	\label{fig:grafo}
	\end{figure}

	We propose a recommendation system, by making matrix completion as in hybrid neighborhood-based collaborative filtering approaches.
	Our approach computes two matrices of similarities, one between users, $S_{\mathcal U}$, and another between items, $S_{\mathcal I}$.
	After, we complete each entry of user $u$ and item $o$ by assigning a convex combination of two quantities, by a parameter $\alpha$.
	The first quantity is a weighted average of the ratings user $u$ gave to other items by the similarities between the other items and item $o$.
	The second quantity is a weighted average of the ratings of item $o$ given by other users similar to user $u$. 
	Figure~\ref{fig:grafo} depicts the users $u_i$ and items $o_j$, connected by an edge with weight $M_{ij}$ whenever user $u_i$ rated item $o_j$. 
	The blue and green edges depict the similarities between users and between items, respectively, with the weights from each similarity matrix $S_{\mathcal U}$ and $S_{\mathcal I}$, respectively.  
		To build the matrices $ S_{\mathcal U}$ and $ S_{\mathcal I}$, we propose two compression similarities based on Kolmogorov complexity, see \cite{cover2012elements}.
		Given the description of a string, $x$, its \emph{Kolmogorov complexity}, $K(x)$, is the length of the smallest computer program that outputs $x$.
		In other words, $K(x)$ is the length of the smallest compressor for $x$. 
		Although Kolmogorov complexity is non-computable, there are efficient and computable approximations by compressors. 
		Let $C$ be a compressor and $C(x)$ denote the length of the output string resulting of the compression of $x$ using $C$.
		The first similarity measure we propose is the following.
		
		\textbf{Compression similarity.}
		 	\emph{Using the \emph{normalized compression distance}, see \cite{li2004similarity}, we define the \emph{compression similarity} as:
				\begin{equation*}
		 			\mathit{CS}( x, y) = 1 - \frac{ C(\tilde x\tilde y) - \min\{C(\tilde x),C(\tilde y)\}}{\max\{C(\tilde x),C(\tilde y)\} },	
		 		\end{equation*}
		 	where string $\tilde x\tilde y$ is the concatenation of $\tilde x$ and $\tilde y$.
			}
		We implement the description of users/items as the string composed by the index of rated items/rating users and respective rating. For instance, if user $u$ rated the items $o_u^1,o_u^2,\hdots,o_u^l$, $l\leq m$, we write the description of user $u$ as the string ``$o_u^1M_{uo_u^1}o_u^2M_{uo_u^2}...o_u^lM_{uo_u^l}$''.
		
		Inspired by CS, in order to reduce the computational complexity, we propose another similarity measure.	
		
		\textbf{Kolmogorov similarity.}
			\emph{We define the \emph{Kolmogorov similarity} as: 
				\begin{equation*}
					\mathit{KS}( x, y)=\left(1+|C(\tilde x)-C(\tilde y)|\right)^{-1}.
				\end{equation*}
			}

		To compress the description strings, we use the standard compression tools from the \emph{zlib} library\footnote{\url{https://tools.ietf.org/html/rfc1950}}.
		Intuitively, both similarities measure how identical are the compactest descriptions of a pair of users or a pair of items.
		
		The compression similarity measures are used to compute the two similarity matrices, $ S_\mathcal U$ and $ S_\mathcal I$.
					
		To complete the rating matrix $M$, we set each non-filled entry $M_{uo}$ in the completed matrix $\hat M$ as a convex combination by parameter $\alpha$ of two quantities. The first is the weighted average of the sum of the ratings of each user $u'\neq u$, weighed by the square of the number of common rated items together with $ S_{\mathcal U_{uu'}}$, $w_o( S_{\mathcal U_{u\cdot}})$.
		The second is the sum of the ratings of each item $o'\neq o$, weighed by the square of the number of user rating the item together with $ S_{\mathcal I_{oo'}}$, $w_u( S_{\mathcal I_{o\cdot}})$.
		Recalling the definitions of $\odot$ and $\|\cdot\|_0$, from Section~\ref{Notation}, the first quantity is given by
		\[
			w_o( S_{\mathcal U_u}) = \frac{1}{z_u}\sum_{u'\neq u} S_{\mathcal U_{uu'}}M_{u'o}\|M_{u\cdot}\odot M_{u'\cdot}\|_0^2, 
		\]
		where
		\[
			z_u = \sum_{u'\neq u} S_{\mathcal U_{uu'}}\|M_{u\cdot}\odot M_{u'\cdot}\|_0^2.
		\]
		Similarly, the second quantity is given by
		\[
			w_u( S_{\mathcal I_o}) = \frac{1}{z_o}\sum_{o'\neq o} S_{\mathcal I_{oo'}}M_{uo'}\|M_{\cdot o}\odot M_{\cdot o'}\|_0^2, 
		\]
		where
		\[
			z_o = \sum_{o'\neq o} S_{\mathcal I_{oo'}}\|M_{\cdot o}\odot M_{\cdot o'}\|_0^2.
		\]
		Lastly, fixed the parameter $0\leq\alpha\leq 1$, we estimate each non filled matrix entry as
		\[
			\hat M_{uo}=\alpha w_o( S_{\mathcal U_{u\cdot}}) + (1-\alpha) w_u( S_{\mathcal I_{o\cdot}}).
		\] 
		Observe that if $\alpha = 1$, it corresponds to user-based collaborative filtering, and if $\alpha = 0$, it corresponds to item-based collaborative filtering.
		The previous steps are summarized in Algorithm~\ref{MatComp}.
		
		Our approach allows to decouple the problem into a set of independent user-by-user subproblems.
		Hence, to generate a set of recommendations for a user, we do not need to complete the entire rating matrix, instead we only need to complete the corresponding matrix row.		
		\begin{algorithm} 
		\caption{Matrix completion algorithm: \textsf{KolMaC}}
		\label{MatComp}
		\begin{algorithmic}[1]
			\STATE{\textbf{input}: $\alpha$, \emph{training set $M$}}
			\STATE{\textbf{compute} $ S_\mathcal U$ from the training set}
			\STATE{\textbf{compute} $ S_\mathcal I$ from the training set}
			\STATE{\textbf{set}} $\hat M = M$
			 \FOR{each user $u$}
			 	\FOR{each item $o$ such that $\hat M_{uo}$ is not filled}
			 			\STATE{\textbf{set} $\hat M_{uo}=\alpha w_o( S_{\mathcal U_{u\cdot}}) + (1-\alpha) w_u( S_{\mathcal I_{o\cdot}})$}
			 	\ENDFOR
			 \ENDFOR
			\STATE{\textbf{output}: $\hat M$}
		\end{algorithmic} 
		\end{algorithm}

	\subsection{Complexity analysis} 
	\label{sub:subsection_name}
	To build the user similarity matrix $ S_\mathcal U$, we first precompute the quantity $C(u)$ for each user $u\in \mathcal U$.
	After, we build an $n\times n$ matrix where each entry $ S_{\mathcal U_{uv}} = \text{KS}(u,v)$ for each $u,v \in \mathcal U$, where we use the pre-computed values from the first step.
	Hence, both time and space complexity for this step are $\mathcal O(n^2)$.
	\emph{Mutatis mutandis}, both time and space complexity to build the item-item similarity matrix $ S_\mathcal I$ are $\mathcal O(m^2)$.
	
	For the similarity measure CS, we perform the same precomputations, but to build matrices $ S_\mathcal U$ and $ S_\mathcal I$, we further need to compute the compression of the concatenation of pairs of users and pairs of products, respectively. Henceforth, the time complexity is $\mathcal O(n^3)$ and $\mathcal O(m^3)$, whilst the space complexity is $\mathcal O(n^2)$ and $\mathcal O(m^2)$, respectively for $ S_\mathcal U$ and $ S_\mathcal I$.
	
	For the matrix completion problem, steps 4-9 of Algorithm~\ref{MatComp}, the time complexity is $\mathcal O(\max\{n,m\})$ (to compute the weighted averages in step 7) times the number of elements of the matrix $n\times m$.
	This yields a time complexity of $\mathcal O(\max\{n^2\times m, n\times m^2\})$.
	The space complexity of those steps is $\mathcal O(n\times m)$.
	
	In summary, the time complexity of Algorithm~\ref{MatComp}, when using KS, is $\mathcal O(\max\{n^2\times m,n\times m^2\})$, and, when using CS, is $\mathcal O(\max\{n^3,m^3\})$. 
	The space complexity of Algorithm~\ref{MatComp} is, for both KS and CS, $\mathcal O(\max\{n^2,m^2\})$.
	
	\section{Experimental setup}\label{Experimentos}
	Next, we describe our experimental settings and analyze the experimental results.
	\subsection{Datasets}
	We test Algorithm~\ref{MatComp} on synthetic and real-world datasets.
	All experiments were done in a 2.8GHz Intel Core 2 Duo, with 4GB 800MHz RAM, using Matlab 2016 and Python 3.
	For the synthetic data, we generate randomly four full rank matrices, with dimension $20\times 30$, and with entries in $\{1,\hdots,5\}$.

	For the real-world datasets we use the MovieLens 100k (ML--100k) and the MovieLens 1M (ML--1M), available in \url{http://movielens.umn.edu}, and both datasets have ratings in $\{1,\hdots,5\}$. Table~\ref{datasets} contain a more detailed description of these datasets.
	
	\begin{table}[ht]
	 	\centering
	 	\begin{tabular}{ | c | c  c  | }
	 		\hline
	 		  & \textsc{ML--100k}    & \textsc{ML--1M}      \\ \hline
			 number of users & 1000			   & 6000				 \\
			 number of items & 	1700	& 4000 \\
			 number of ratings &	100,000		&1,000,000\\	 	\hline
	 	\end{tabular}
	 	\caption{RMSE for the datasets ML--100k and ML--1M.}
	 	\label{datasets}
	 \end{table}%

		\subsection{Evaluation metric}
	To evaluate and compare the performance of the proposed algorithm, Algorithm~\ref{MatComp}, we use the 5-fold-cross-validation method on both synthetic and real data. For the ML--100k, the dataset already provides a set of $5$ train and test files. For the ML--1M we randomly split the original dataset in a set of $5$ train/test files.
	In the synthetic data the four randomly generated full rank matrices, with dimension $20\times 30$, were split as in the ML--1M case.

	 We use the \emph{root-mean-square error (RMSE)} \cite{koren2008factorization} to evaluate the performance of the proposed algorithm by measuring the difference between the estimated missing values and the original values.
	 Let $M$ be the original matrix, $M^\ast$ equal to $M$ except on the missing entries of the test set $\mathcal T$, and let $\hat M$ be the estimation of $M$ by a matrix completion method when applied to $M^\ast$.
	 The RMSE is given by
	 \[
	\text{RMSE}(M,\hat M)=\sqrt{\frac{1}{|\mathcal T|}\sum_{(i,j)\in\mathcal T}(M_{ij}-\hat M_{ij})^2}.
	 \]
		
	\subsection{Experimental results}	
		We use the above described datasets to test our algorithm, using both similarity measures KS and CS, against the following algorithms:
		NormalPredictor, BaselineOnly \cite{koren2010factor}, KNNBasic \cite{altman1992introduction}, KNNWithMeans \cite{altman1992introduction}, KNNBaseline \cite{koren2010factor}, SVD \cite{salakhutdinov2007probabilistic}, SVD++ \cite{koren2008factorization}, NMF \cite{lee2001algorithms}, Slope One \cite{lemire2005slope} and Co-clustering \cite{george2005scalable}.
		This set of algorithms is implemented in the Python toolkit Surprise\footnote{\url{http://surpriselib.com/}}. 
		The results of the experiments are summarized in Table~\ref{tab:synth}, for the synthetic data, and in Table~\ref{tab:real}, for the real datasets.
		For the synthetic data, the best result corresponds to using Algorithm~\ref{MatComp}, with the similarity CS.
		When using similarity KS, the result is the third best in the set of tested methods.
		This happens because the majority of the compared methods assume that the matrix they are completing is low rank, which might be the case in these datasets, but might not be the case in general.
		
		With real data, using both KS and CS similarity measures, our algorithm does not have the lowest RMSE, which may happen due to the fact that most of the compared methods assume the completed matrix is low rank.
		However, the results are comparable and of the same order as the best reported ones. 
		The advantages of our algorithm are: it can be computed in a distributed fashion, does not need assumptions on the matrix rank, does not need to known the dimensions of the subspaces neither initialization, does not estimate latent variables, and it is model free.
		Finally, it scales better than the methods with better RMSE, on the real data, than our method. 
	\begin{table}[ht]
	 	\centering
	 	\begin{tabular}{ | c | c c c c | }
	 		\hline
	 		 \textsc{Method} &  $M_1$  & $M_2$ & $M_3$ & $M_4$    \\ \hline\hline
			 NormalPredictor & 1.8692 & 1.8944 & 1.7140 & 1.9263\\
			 BaselineOnly & 1.4667 & 1.4663 & 1.4306 & 1.4803\\
			 KNNBasic &	1.4665 & 1.4840 & 1.4383 & 1.5049\\
			 KNNWithMeans &	1.5107 & 1.5150 & 1.4721 & 1.5269\\
			 KNNBaseline &	1.4838 & 1.4998 & 1.4549 & 1.5126\\
			 SVD &	1.5155 & 1.5120 & 1.4660 & 1.5222\\
			 SVD++ &	1.5205 & 1.5176 & 1.4698 & 1.5279\\
			 NMF &	1.6999 & 1.6703 & 1.7052 & 1.7686\\
			 Slope One &	1.5270 & 1.5287 & 1.4760 & 1.5310\\
			 Co-clustering &	1.5808 & 1.5630 & 1.5461 & 1.6442\\ \hline
			 \textbf{KolMaC KS}& 1.4689	& 1.4676  & 1.4303 & 1.4848  \\
			 \textbf{KolMaC CS}& \textbf{1.4530} &	\textbf{1.4520} & \textbf{1.4260} & \textbf{1.4714}   \\ 
	 	\hline
	 	\end{tabular}
	 	\caption{RMSE of a 5-fold-cross-validation in four synthetic random and full rank $20\times 30$ matrices.}
	 	\label{tab:synth}
	 \end{table}%


	\begin{table}[ht]
	 	\centering
	 	\begin{tabular}{ | c | c  c  | }
	 		\hline
	 		 \textsc{Method} & \textsc{ML--100k}    & \textsc{ML--1M}      \\ \hline\hline
			 NormalPredictor & 1.5228			   & 1.5037				 \\
			 BaselineOnly & 	0.9445	& 0.9086 \\
			 KNNBasic &	0.9789		&0.9207\\
			 KNNWithMeans &	0.9514	& 0.9292\\
			 KNNBaseline &	0.9306& 0.8949\\
			 SVD &	0.9396	& \textbf{0.8936}\\
			 SVD++ &	\textbf{0.9200}&	--\\
			 NMF &	0.9634	&  0.9155\\
			 Slope One &	0.9454&	 0.9065\\
			 Co-clustering &	0.9678&	 0.9155\\ 
			 \hline
			 \textbf{KolMaC KS}			   &     0.9660 & 0.9330
 \\
			 \textbf{KolMaC CS} 		 &   0.9618
 & \\  
	 	\hline
	 	\end{tabular}
	 	\caption{RMSE for the datasets ML--100k and ML--1M.}
	 	\label{tab:real}
	 \end{table}%
		
\section{Conclusions}\label{Conclusion}
	We present a novel hybrid neighborhood-based collaborative filtering recommendation system, by making independent user-by-user matrix completion, that uses Kolmogorov complexity.
	Our method does not require assumptions about the rank of the matrix, does not need to specify dimensions of subspaces, it is model free, and therefore it is more general.
	We present experimental results on both synthetic and real dataset which show that our approach is comparable with state of the art approaches.
	The avenues for further research include exploring matrix completion with the presence of noise, and to extend this work, where in an initial step, we cluster users and items by using the similarities between users and items, respectively.

\bibliographystyle{apalike}
\bibliography{library.bib}

\end{document}